# Fractional charges and Misner-Wheeler "charge without charge" effect in metamaterials

Igor I. Smolyaninov

*Department of Electrical and Computer Engineering, University of Maryland, College Park, MD 20742, USA*

**Optical space in metamaterials may be engineered to emulate four dimensional Kaluza-Klein theory. Nonlinear optics of such metamaterials mimics interaction of quantized electric charges. An electromagnetic wormhole is designed, which connects two points of such an optical space and changes its effective topology. Electromagnetic field configurations which exhibit fractional charges appear as a result of such topology change. Moreover, such effects as Misner-Wheeler "charge without charge" may be replicated.**

Recent development of artificial electromagnetic metamaterials enables unprecedented control over the coordinate dependencies of the dielectric permittivity $\varepsilon_{ik}$ and magnetic permeability $\mu_{ik}$ tensors of an electromagnetic medium. This development gave rise to numerous novel device ideas based on the concept of "electromagnetic space", which is different from the actual physical space [1-4], and may have non-trivial topology [5,6]. While current emphasis of research in this field is concentrated in the area of novel linear electromagnetic devices, nonlinear optics of metamaterials also appears to be extremely interesting. For example, nonlinear optics of metamaterials has a unique capability to realize table top models of many gravitational and field theoretical phenomena [7-9]. In particular, very recently it was demonstrated that using extraordinary waves in anisotropic uniaxial metamaterials a model of four-dimensional Kaluza-Klein theory may be created [7]. Nonlinear optics of such a metamaterial space was shown to resemble interaction of effective electric charges. Here we demonstrate that similar to recent reports in refs.[5,6], an electromagnetic wormhole may be designed, which connects two points of such an optical space and changes its effective topology. Electromagnetic field configurations which exhibit fractional charges may appear as a result of such topology change.

Let us start with a brief review of four dimensional Kaluza-Klein model, which is based on electromagnetic metamaterials. Spatial geometry of this space-time may be approximated as a product $R_2 \times S_1$ of a 2D plane $R_2$ and a circle $S_1$, as shown in Fig.1(a). Its line element may be written as

$$dl^2 = dx^2 + dy^2 + R^2 d\phi^2 \tag{1}$$

Using the stereographic projection $z=2R\sin\phi/(1+\cos\phi)$, this line element may be re-written as

$$dl^2 = dx^2 + dy^2 + \frac{dz^2}{\left(1 + \frac{z^2}{4R^2}\right)^2} \tag{2}$$

Equation (2) indicates that we need a uniaxial anisotropic metamaterial in order to emulate the $R_2 \times S_1$ space. Let us consider a non-magnetic uniaxial anisotropic metamaterial with dielectric permittivities $\varepsilon_x = \varepsilon_y = \varepsilon_1$ and $\varepsilon_z = \varepsilon_2$. The wave equation in such a material may be written as

$$-\frac{\partial^2 \vec{E}}{c^2 \partial t^2} = \vec{\vec{\varepsilon}}^{-1} \vec{\nabla} \times \vec{\nabla} \times \vec{E} \qquad (3)$$

where $\vec{\vec{\varepsilon}}^{-1} = \vec{\vec{\xi}}$ is the inverse dielectric permittivity tensor calculated at the center frequency $\omega$ of the signal bandwidth [10]. Any electromagnetic field propagating in this material may be expressed as a sum of the "ordinary" ($\vec{E}$ perpendicular to the optical axis) and "extraordinary" ($\vec{E}$ parallel to the plane defined by the k–vector of the wave and the optical axis) contributions. Let us define an extraordinary wave function as $\psi = E_z$ (so that the ordinary portion of the electromagnetic field does not contribute to $\psi$). Equation (3) then yields:

$$\frac{\partial^2 \psi}{c^2 \partial t^2} = \frac{\partial^2 \psi}{\varepsilon_1 \partial z^2} + \frac{1}{\varepsilon_2}\left(\frac{\partial^2 \psi}{\partial x^2} + \frac{\partial^2 \psi}{\partial y^2}\right) \qquad (4)$$

Extraordinary field will perceive the optical space as $R_2 \times S_1$ if

$$\varepsilon_2 = n^2 \text{ and } \varepsilon_1 = \frac{n^2}{\left(1 + \frac{z^2}{4R^2}\right)^2} \qquad (5)$$

where $n$ is a constant. Such an anisotropic uniaxial metamaterial may be realized using a 3D layered hyperbolic metamaterial structure shown in Fig.1(b). Let us assume that the metallic layers are oriented perpendicular to the $z$ direction. The diagonal components of the permittivity tensor in this case have been calculated in ref. [11] using Maxwell-Garnett approximation:

$$\varepsilon_1 = \alpha \varepsilon_m + (1-\alpha)\varepsilon_d \;, \quad \varepsilon_2 = \frac{\varepsilon_m \varepsilon_d}{(1-\alpha)\varepsilon_m + \alpha \varepsilon_d} \qquad (6)$$

where $\alpha$ is the fraction of metallic phase, and $\varepsilon_m<0$ and $\varepsilon_d>0$ are the dielectric permittivities of metal and dielectric layers, respectively. We would like to produce the anisotropic dielectric permittivity behavior described by eq.(5) by changing $\alpha$ as a function of $z$. Simple analysis of eqs.(5,6) indicates that

$$\alpha = \frac{\varepsilon_d - n^2/\left(1+\frac{z^2}{4R^2}\right)^2}{\varepsilon_d - \varepsilon_m} \tag{7}$$

produces the required $z$ dependence of $\varepsilon_1$, while $\varepsilon_2$ remains approximately constant if $-\varepsilon_m >> \varepsilon_d$.

Let us consider nonlinear optics of the $R_2 \times S_1$ metamaterial space, and demonstrate that it resembles the picture of effective charges interacting with each other via gauge fields. This result is natural, since nonlinear optics of this space is modelled after the Kaluza-Klein theory. The eigenmodes of the extraordinary field may be written as

$$\psi_{kL} = e^{ik\rho} e^{iL\phi} \tag{8}$$

where $L$ is the quantized integer "angular momentum" number and $k$ is the 2D momentum. The dispersion law of these eigenmodes is

$$\frac{n^2}{c^2}\omega^2 = k^2 + \frac{L^2}{R^2} \tag{9}$$

Therefore, extraordinary photons having $L=0$ behave as massless 2D quasiparticles, while $L \neq 0$ photons are massive. Let us demonstrate that the "angular momentum" number $L$ behaves as a conserved quantized integer effective charge in the nonlinear optical interactions of extraordinary photons.

Nonlinear optical effects deform $\vec{\vec{\varepsilon}}^{-1} = \vec{\vec{\xi}}$ tensor and therefore deform the line element (1). In the weak field approximation corrections to $\vec{\vec{\varepsilon}}^{-1}$ are small. However, corrections to the off-diagonal terms of $\vec{\vec{\varepsilon}}^{-1}$ cannot be neglected. Therefore, the effective metric of the deformed optical space may be written as

$$ds^2 = g_{\alpha\beta}dx^\alpha dx^\beta + 2g_{\alpha 3}dx^\alpha d\phi + g_{33}d\phi^2 \qquad (10)$$

where the Greek indices $\alpha$=0, 1, 2 indicate coordinates of the (almost flat) planar 3D Minkowski space-time: $dx^0=cdt$, $dx^1=dx$, and $dx^2=dy$. This 3D space-time is populated by extraordinary photons (described by "scalar" wave function $\psi$) which are affected by the vector field $g_{\alpha 3}$ (its components are $g_{03}$=0, $g_{13}=2\xi_{13}d\phi/dz$, and $g_{23}=2\xi_{23}d\phi/dz$) and the scalar field $g_{33}=R^2$, which is usually called the dilaton. The dilaton field may be assumed constant in the weak field approximation. For a given value of $L$, the wave equation may be written as

$$\hat{D}\psi = \Box\psi - L^2\frac{1-g_{\alpha 3}g^{\alpha 3}}{g_{33}}\psi + 2iLg^{\alpha 3}\frac{\partial\psi}{\partial x^\alpha} + iL\frac{\partial g^{\alpha 3}}{\partial x^\alpha}\psi = 0 \qquad (11)$$

where $\Box$ is the covariant three-dimensional d'Alembert operator. Equation (11) looks exactly the same as the Klein-Gordon equation for a charged particle. In 3D space-time it describes a particle of mass

$$m = \frac{\hbar L}{cg_{33}^{1/2}} \qquad (12)$$

which interacts with a vector field $g^{\alpha 3}$ (playing the role of a vector potential) via its quantized charge $L$. The linear portion of $\hat{D}$ in equation (11) describes standard metamaterial optics in which the metamaterial plays the role of a curvilinear background metric. On the other hand, the third order optical nonlinearity of the form

$$g^{\alpha 3} \propto S_3 \propto (E_1 B_2 - E_2 B_1) \propto L, \tag{13}$$

where $S_3$ is the $z$ component of the Poynting vector, leads to Coulomb-like interaction of the effective charges with each other. Extraordinary photons having nonzero angular momentum $L \neq 0$ act as sources of the $g^{\alpha 3}$ field, which in turn acts on other "charged" ($L \neq 0$) extraordinary photons. Such a nonlinear interaction may be realized using a dielectric metamaterial component $\varepsilon_d$ (see eq.(6)), which exhibits recently suggested Poynting nonlinearity in nonlinear Fabry-Perot (NLFP) resonator geometry [12]. The proposed metamaterial geometry shown in Fig.1(b) is indeed quite similar to the NLFP geometry.

Since $g^{03}=0$ (which corresponds to Weil or temporal gauge), the effective "electric field" in our model equals

$$\vec{E}_{eff} = \left(i\omega g^{13}, i\omega g^{23}\right) = \nabla f_{eff} \tag{14}$$

Therefore, a natural choice of the effective "potential" $f_{eff}$ is

$$f_{eff} = (E_1 B_2 - E_2 B_1) \tag{15}$$

Indeed, such a choice leads to effective "Poisson equation"

$$\Delta_2 f_{eff} = -4k^2 (E_1 B_2 - E_2 B_1) \propto L, \tag{16}$$

where $\Delta_2 = \partial^2/\partial x^2 + \partial^2/\partial y^2$ and $k$ is defined by eq.(9). By its definition as the "angular momentum" number, such an effective charge $L$ is conserved. At large enough R and small L

$$k \approx \frac{n}{c}\omega \tag{17}$$

The effect of such 2D effective Coulomb interaction cannot be neglected when the kinetic energy term in eq.(11) becomes comparable with the potential energy terms.

As a second step, let us demonstrate that similar to the recent reports in refs.[5,6], an electromagnetic wormhole may be designed, which connects two points of the $R_2xS_1$ metamaterial space and changes its effective topology. Based on the spatial distribution of the dielectric permittivity tensor components given by eq.(5), it is clear that the electromagnetic field distribution in the $R_2xS_1$ metamaterial space must be somewhat similar to a field distribution in a planar waveguide. This conclusion has been confirmed by numerical calculations using Comsol Multiphysics 4.2a solver, as shown in Fig.1(c). Therefore, design of an electromagnetic wormhole connecting two points of the $R_2xS_1$ metamaterial space appears to be similar to the design of a plasmonic analogue of an electromagnetic wormhole described in [6]. Such a wormhole may be designed as a toroidal handlebody (Fig.2(a)), which bridges two remote locations of the planar waveguide. We may implement the same recipe as in [6], where the metamaterial parameters for an invisible toroidal handlebody were designed as follows. Using toroidal coordinates *(r,u,v)*, which are related to the Cartesian coordinates *(x,y,z)* as

$$x = r\cos u \qquad (18)$$

$$y = (R + r\sin u)\sin v$$

$$z = (R + r\sin u)\cos v$$

(where *2R* is the distance between points connected by the wormhole), the metamaterial parameters of the handlebody may be obtained by coordinate transformations

$$r' = a + (b-a)r/b, \; u'=u, \; v'=v, \qquad (19)$$

where *a* and *b* are the inner and outer radii of the cloaked wormhole area, respectively. The corresponding material parameters are [6]:

$$\varepsilon_{rr} = \mu_{rr} = \frac{r-a}{r}\frac{(b-a)R+b(r-a)\sin u}{(b-a)(R+r\sin u)} \quad (20)$$

$$\varepsilon_{uu} = \mu_{uu} = \frac{r}{r-a}\frac{(b-a)R+b(r-a)\sin u}{(b-a)(R+r\sin u)}$$

$$\varepsilon_{vv} = \mu_{vv} = \frac{b^2}{b-a}\frac{r-a}{r}\frac{R+r\sin u}{(b-a)R+b(r-a)\sin u}$$

However unlike the invisible "plasmonic electromagnetic wormhole" prescription in [6], we do not need to make the handlebody invisible for light waves incident on the plasmonic waveguide from the outside 3D space. There are no such "outside" waves in the $R_2xS_1$ metamaterial space. Moreover, as clearly demonstrated by Fig.1(c), all waves which exist in the $R_2xS_1$ space attenuate strongly away from the z=0 plane. Since our goal is only to change the effective topology of the metamaterial space, the requirement of wormhole "invisibility" does not need to be strictly observed. As a result, complicated spatial material parameters distribution described by eqs.(20) may be simplified without considerable effect on visibility of the toroidal handlebody. Such a "reduced visibility" wormhole may be implemented using a toroidal handlebody having isotropic homogeneous $\varepsilon$. Numerical simulations of such a wormhole are shown in Figs.2(b,c).

Based on the "effective Gauss theorem" expressed by eqs.(14-16), it is clear that wormhole-induced topology changes of the $R_2xS_1$ metamaterial space must result in electromagnetic field configurations which exhibit fractional effective charges. Indeed, we may introduce an effective "electric field flux" through a closed contour C enclosing an area A of the $R_2$ plane as

$$\Phi_{Eeff} = \oint_C \vec{E}_{eff} d\vec{l} , \qquad (21)$$

where $d\vec{l}$ is a vector representing an infinitesimal element of the contour length oriented normally to the contour. According to eq.(16), such an effective flux is quantized if the metamaterial $R_2 \times S_1$ space is simply connected. On the other hand, if a wormhole connects an inside point of the area A to some other distant point (as shown in Fig.2(a)), the integral in eq.(21) may produce a non-integer fractional value. This effect may be illustrated by Fig.3, which shows numerical simulations of power flow near the electromagnetic wormhole in the same geometry as in Fig.2(b,c). The average Poynting vector directions are indicated by black arrows. Near the wormhole openings the *z* components of the Poynting vector $S_3$ are nonzero and have opposite signs, which according to eq.(13-16) leads to appearance of opposite contributions to the effective charge of the wormhole openings. Appearance of fractional charges in our metamaterial model is interesting because such effects are relatively rare. Fractional charges are known to appear in quantum chromodynamics as fractional charges of quarks [13], and in systems, which exhibit the fractional quantum Hall effect [14]. Studying such metamaterial space configurations also appears interesting in light of recent demonstrations that two colored quasiparticles (quarks) in maximally supersymmetric Yang-Mills theory entangled in a color singlet EPR pair are connected by a wormhole [15,16], and that the physical vacuum appears to exhibit hyperbolic metamaterial properties when subjected to a very strong magnetic field [17]. In addition, such effects as Misner-Wheeler "charge without charge" [18] and the recently proposed charge-hiding effect [19] may be replicated in the metamaterial system. Indeed, as obvious from Fig.3, coupling of some of the *L*=0 mode power into the wormhole leads to appearance of opposite effective charges of the wormhole openings due to non-zero $S_3$ near the openings.

As far as the fractional charge values are concerned, any non-integer value of the effective charge may appear in our model if the wormhole geometry is arbitrary. However, if the wormhole geometry exhibits some symmetry (such as symmetry under rotation by either $\pi$ or $2\pi/3$ radian, as shown schematically in Figs.4(a,b)), the allowed values of effective charges may become restricted to some simple fractions. The total charge of a closed area $A$ defined by eq.(21) must be integer if all the wormholes, which originate inside $A$ do not lead outside of this area. If a rotational symmetry is imposed on the electromagnetic field configurations, the integer effective charge inside $A$ must be divided equally between the wormhole openings, leading to appearance of either 1/2 or 1/3 charges in wormhole configurations shown schematically in Figs.4(a,b). Because of the effective charge conservation, a photon with $L=-1$ traveling through one of the openings of the latter wormhole may also lead to appearance of -2/3 charge.

In conclusion, we have demonstrated that optics of metamaterials presents us with new opportunities to engineer topologically non-trivial "optical spaces". Nonlinear optics of extraordinary light in these spaces resembles Coulomb interaction of effective charges. Electromagnetic wormholes may be designed which connect two points of such an optical space and change its effective topology. Electromagnetic field configurations which exhibit fractional charges appear as a result of such topology change. Moreover, such effects as Misner-Wheeler "charge without charge" [18] may be replicated.

**Figure captions**

**Fig.1** (a) "Optical space" in an anisotropic uniaxial metamaterial may mimic such topologically non-trivial 3D space as $R_2 \times S_1$, which is a product of a 2D plane $R_2$ and a circle $S_1$. (b) Schematic view of the "layered" 3D hyperbolic metamaterial made of subwavelength metal and dielectric layers, which can be used to emulate the $R_2 \times S_1$ space. (c) Example of numerical calculations of electromagnetic field distribution in the *xz* plane inside the $R_2 \times S_1$ metamaterial space using Comsol Multiphysics 4.2a solver. The calculated power flow distribution is similar to power flow in a multimode planar waveguide.

**Fig.2** (a) Schematic view of a toroidal handlebody, which connects two points of the $R_2 \times S_1$ metamaterial space and changes its effective topology. (b,c) Numerical simulations of the electromagnetic wormhole: (b) spatial distribution of $\varepsilon_l$, and (c) corresponding spatial distribution of $B_y$.

**Fig.3** Numerical simulations of power flow near the electromagnetic wormhole in the same geometry as in Fig.2(b,c), which demonstrates "charge without charge" effect. The average Poynting vector direction is indicated by black arrows. Near the wormhole openings the *z* components of the Poynting vector $S_3$ are nonzero and have opposite signs, which according to eq.(11-14) leads to appearance of opposite contributions to the effective charge of the wormhole openings (the openings are marked by dashed lines).

**Fig.4** If the wormhole geometry exhibits some symmetry (such as symmetry under rotation by either $\pi$ (a) or $2\pi/3$ (b) radian, the allowed values of effective charges of the wormhole openings may become restricted to some simple fractions of the "elementary charge" *e*: if such a rotational symmetry is imposed on electromagnetic field

configurations, the integer effective charge inside *A* must be divided equally between the wormhole openings.

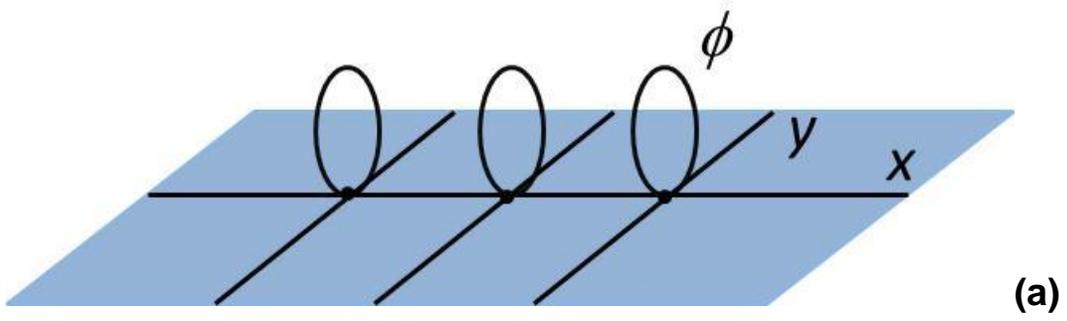

(a)

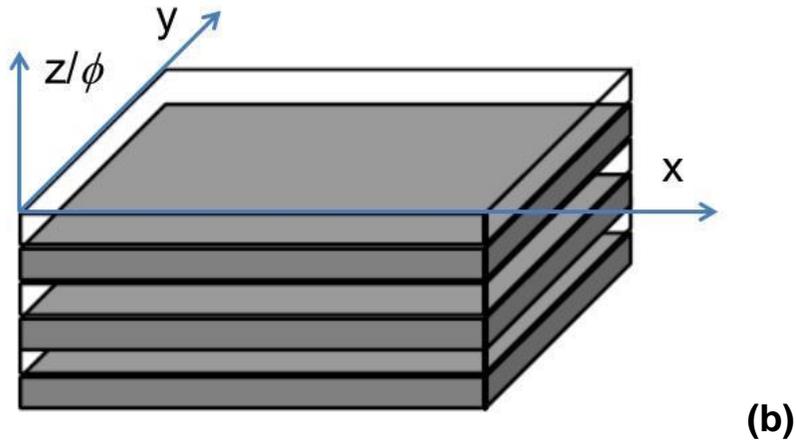

(b)

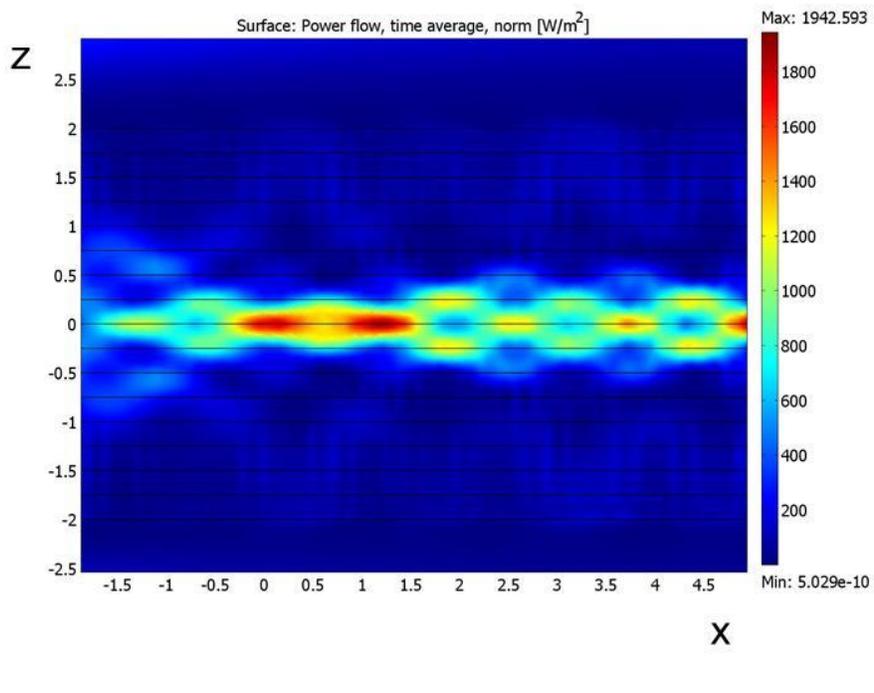

(c)

Fig.1

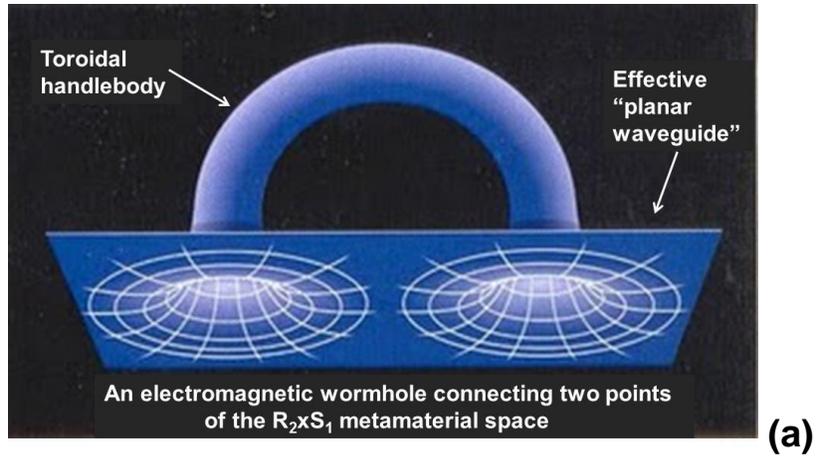
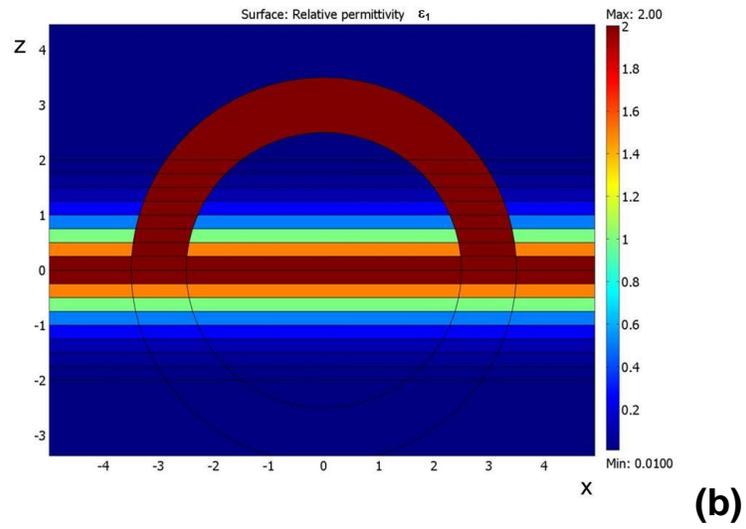
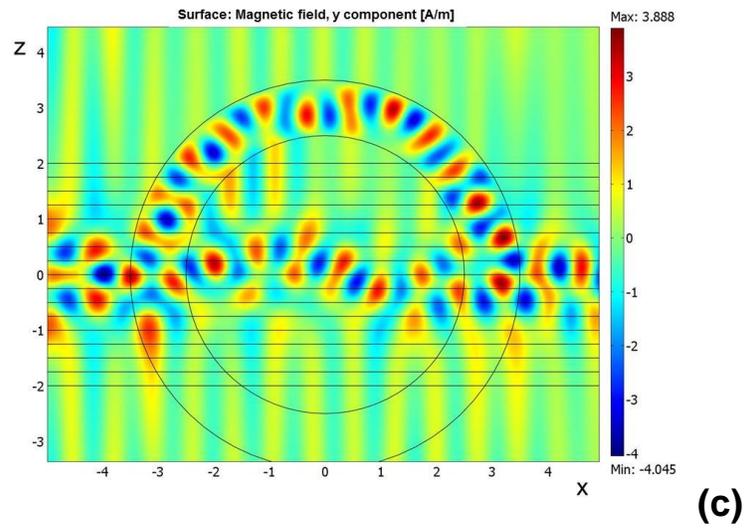

Fig. 2

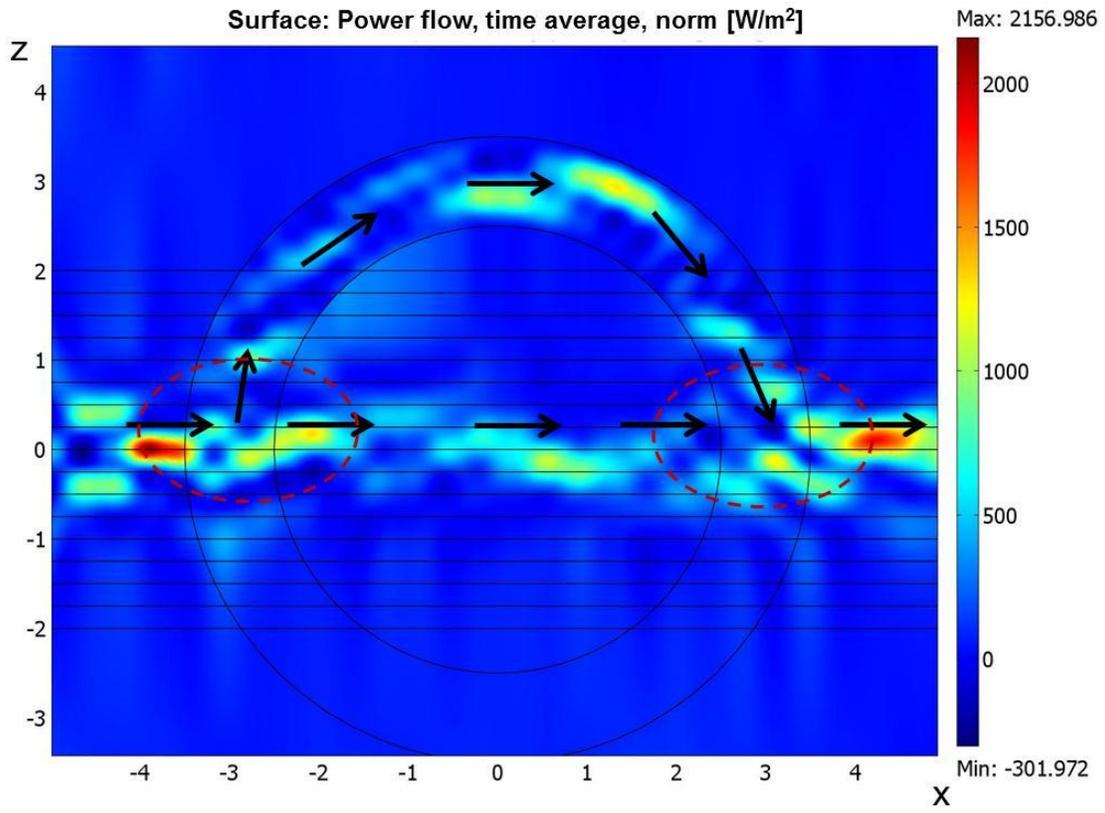

Fig. 3

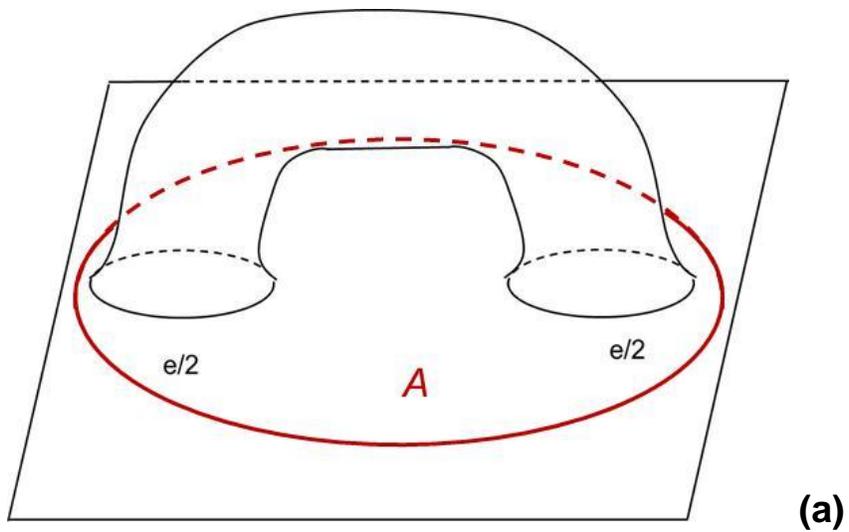

**(a)**

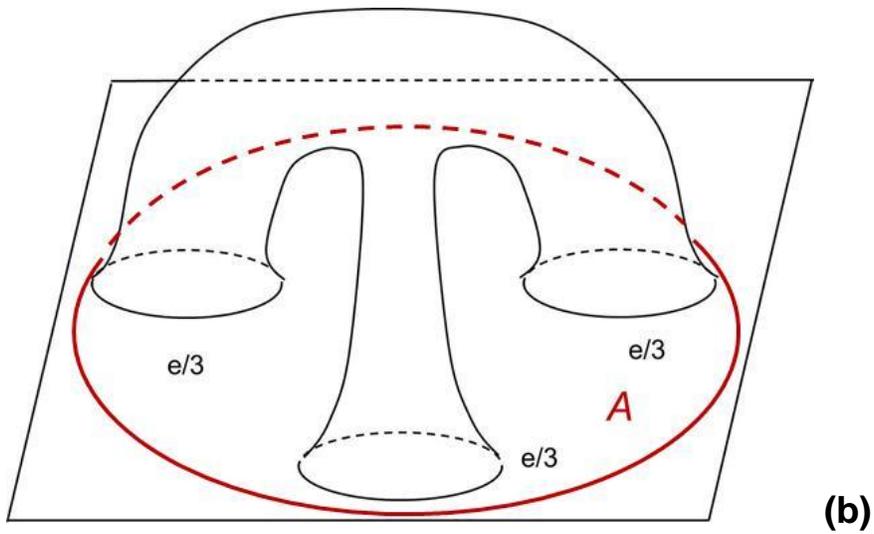

**(b)**

Fig. 4